# Analyzing Counterintuitive Data

Erik Doty[1], DO, MBI; Ned McCague[2,5], MPH; David J. Stone[3,5], MD; Leo Anthony Celi[4,5], MD, MS, MPH

Corresponding author: Erik Doty

Email: edoty@une.edu

Phone: 603-718-9809

1. *Baystate Medical Center, Springfield, MA*
2. *Kyruus, Inc., Boston, MA*
3. *University of Virginia School of Medicine, Charlottesville, VA*
4. *Beth Israel Deaconess Medical Center, Boston, MA*
5. *MIT Laboratory for Computational Physiology, Cambridge, MA*

# Analyzing counterintuitive data

E. Doty, N. McCague, D.J. Stone, L.A. Celi

## Abstract

**Purpose:** To explore the issue of counterintuitive data via analysis of a representative case and further discussion of those situations in which the data appear to be inconsistent with current knowledge.

**Case:** 844 postoperative CABG patients, who were extubated within 24 hours of surgery were identified in a critical care database (MIMIC-III). Nurse elicited pain scores were documented throughout their hospital stay on a scale of 0 to 10. Levels were tracked as mean, median, and maximum values, and categorized as no (0/10), mild (1-3), moderate (4-6) and severe pain (7-10). Regression analysis was employed to analyze the relationship between pain scores and outcomes of interest (mortality and hospital LOS). After covariate adjustment, increased levels of pain were found to be associated with lower mortality rates and reduced hospital LOS.

**Conclusion:** These counterintuitive results for post-CABG pain related outcomes have not been previously reported. While not representing strong enough evidence to alter clinical practice, confirmed and reliable results such as these should serve as a research trigger and prompt further studies into unexpected associations between pain and patient outcomes. With the advent of frequent secondary analysis of electronic health records, such counterintuitive data results are likely to become more frequent. We discuss the issue of counterintuitive data in extended fashion, including possible reasons for, and approaches to, this phenomenon.

## Introduction

What do we mean by counterintuitive data?  It is data that presents unexpected results that may clash with common sense or what has been previously published and accepted by the medical community. In practice, clinicians have long dealt with such results in individual bits but have had the vast advantage of being able to examine the concurrent state of the patient and react in real time by repeating a lab test or tracking ongoing monitor data. These responses  function to identify the prior result as a non-repeatable error, or as a genuine anomaly. However, this approach is not applicable to the context of retrospective data analysis. Furthermore, the counterintuitive data revealed in such analyses is likely to be more involved than a single aberrant lab or vital sign value. In today's data driven healthcare system, retrospective data analyses are becoming more and more common. We can therefore logically expect to encounter a greater incidence and variety of counterintuitive values and results that are impossible to confirm by repetition, difficult to confirm or deny by context, but still require interpretation.

The question then becomes how best to approach such results? Are they incorrect simply because they weren't what was expected? And was the expectation itself based on subjective assumptions or objective conclusions? When our prior expectations are not met, are we dealing with truly faulty data, or do our expectations need to be reset by what are actually reliable, but counterintuitive, results. For example, we have learned that intensive care practices common in the past such as large tidal volume ventilation, the use of pulmonary artery catheters, and the use of lidocaine infusions in myocardial infarction led to no benefit or injury.[1-3] Were these unexpected negative outcomes initially missed because outcomes data was not being carefully analyzed, or perhaps ignored or interpreted as counterintuitive to the level of unbelievability? How can the situation be dissected retrospectively so that counterintuitive data can be identified as truly spurious versus simply not being consistent with our prior experience which may itself be faulty and require data driven correction?

In this paper, we explore a case in which the results contradicted previous reports and our clinical expectations. Using the Medical Information Mart for Intensive Care-III (MIMIC-III), a critical care database that was developed and maintained by the Laboratory for Computational Physiology at the Massachusetts Institute of Technology[4], we retrospectively selected a cohort of patients that underwent a coronary artery bypass graft (CABG) procedure and evaluated the effect of perceived pain on mortality and hospital length of stay (LOS). We found that higher levels of pain were associated with reduced mortality and reduced LOS, contrary to previous reports in which better pain control was associated with better outcomes. We then discuss potential causes of these results and the general issue of dealing with counterintuitive results in retrospective data analyses.

## Case

### Population

We selected patients from the MIMIC database who met all of the following inclusion criteria and none of the exclusion criteria. Inclusion criteria included: (1) Adult > 18 years old, (2) who underwent CABG surgery, and (3) was extubated within 24 hours after arrival to the ICU. Exclusion criteria were: (1) Non-CABG surgical procedure, and (2) missing data on confounding variables. Patients were identified using Current Procedural Terminology (CPT) codes: The following CPT codes corresponded to the CABG procedure: 33510 to 33516 for venous grafting only for coronary artery bypass, and 33533 to 33548 for arterial grafting for coronary bypass. The final study cohort contained 844 patients (*Figure 1*).

The MIMC-III database included 1,917 patients who underwent CABG, with 844 meeting the study criteria. CABG was chosen for the investigation as it is a common procedure with the majority of patients having no or few post-operative complications and relatively predictable recoveries.[5] Due to the nature of the surgical procedure which requires sternal spreading for exposure, there is an expected high analgesic burden immediately after surgery.

## Outcomes

The primary outcome assessed was mortality at 30 days. Secondary outcomes were mortality at 1 year and hospital LOS. In the MIMIC database, mortality data for patients who die after hospital discharge is derived from the social security death registry.[4]

## Exposures

The exposures of interest were pain levels reported by the patient immediately and in the subsequent interval after ICU extubation. Pain levels on a scale of 0-10 were regularly self-reported by patients to ICU nurses and recorded in the database, generating a continuum of measurements for each patient. The mean, median, and maximum pain levels were used for separate analyses. Concomitant measurements of heart rates, respiratory rates, and systolic blood pressures were also compared against their simultaneously recorded pain measurement. Nausea and delirium were also tested against our outcomes. The presence of nausea was derived from the nursing notes stored in the database. A positive nausea exposure was defined as the mention "nausea" or "nauseous" in the nursing note with no negative descriptor, such as "not nauseous" or "denies nausea", attached. Delirium was similarly assessed by looking for mention of "delirium", "delirious", or "confusion". Additionally, delirium exposure was considered positive if patients had a positive nursing delirium assessment.

## Covariates

Several variables were included to control for confounding in the regression models: demographic factors, comorbid conditions, and illness severity score on admission to the ICU.[6] Comorbid burden was represented by the Elixhauser index which is determined by the aggregate presence or absence of 30 different comorbid conditions as detected by ICD-9 codes.[7] Illness severity was captured using the Oxford Acute Severity of Illness Score (OASIS), which is calculated on admission to the ICU and takes into account age, heart rate, Glasgow coma scale, mean arterial pressure, temperature, respiratory rate, ventilatory status, urine output, pre-ICU in-hospital LOS, and whether or not the patient underwent elective surgery. Studies have shown OASIS is comparable to other illness severity ratings in predicting outcomes such as mortality and length of stay.[8]

## Analysis

Analysis was carried out using R version 3.4.0 and SAS 9.4. Unconditional logistic regression with Fisher's optimization was used to compare the pain measures with 30-day and 1-year mortality. Linear regression was used to model the relationship between mean pain scores and hospital LOS. Age, gender (male reference), Elixhauser index, and OASIS score were included in the models to account for potential confounders. In a separate ordinal regression, mean pain levels were categorized into four groups of no pain (0/10), mild pain (1-3), moderate pain (3-6), and severe pain (7-10) in accordance with the NIH Pain Consortium.[9] ANOVA was used to determine if there was a significant variation in heart rate, respiratory rate, and/or systolic blood pressure, when compared to the concurrent pain assessment. Two sensitivity analyses were performed to assess the robustness of the observed effects. The first included the same statistical tests in all postoperative CABG patients regardless of duration of intubation. The second sensitivity analysis excluded patients who died in the hospital. To add validity to the potential associations, falsification hypothesis testing using nausea, a symptom with no known effect on clinical outcomes, was performed on the same patient cohort. Assessment of delirium, a symptom associated with poorer patient outcomes, was also performed against the outcome measures.[10]

## Results

The database included 844 patients who underwent a CABG procedure and were extubated within 24 hours. There were 68 patients who on average reported no pain during their ICU stay after extubation, 419 with mild pain, 336 with moderate pain, and 21 with severe pain. The distribution of patient characteristics, including age, gender, illness acuity on ICU admission (OASIS), and comorbidity index is reported in *Table 1*. There was no significant difference noted in the frequency in which pain was assessed in those who experienced lower pain levels when compared to those who experienced increased pain levels. The number of comorbidities ranged from 0 to 9. Bivariate analysis showed increasing OASIS was significantly associated with increased mortality and increased LOS ($p < 0.05$).

Bivariate analysis (*Figure 2*) shows a correlation between increasing pain levels and improved outcomes among these patients who had no intra-operative complications and were extubated within 24 hours of arrival in the ICU. Higher pain levels for this specific cohort of patients who were fast-tracked after CABG were found to be associated with decreased hospital LOS. Those who experienced lower levels of pain in the ICU were more likely to be dead at 30 days and 1 year.

Multivariate regression analysis was performed to adjust for confounding. Four different models using mean, median, and maximum pain scores, and pain categories were tested against the clinical outcomes with the results displayed in *Table 2*. The logistic regression models consistently showed that increasing pain was associated with reduced odds of death at 30 days and 1 year after adjustment for illness severity and co-morbid conditions. All the linear models demonstrated that increasing pain levels were also associated with decreased hospital LOS, except for the model that looked at the maximum pain score, which showed an opposite effect.

No significant variations were noted in heart rate, respiratory rate, or blood pressure with increasing pain levels.

Sensitivity analysis was employed to examine all patients regardless of duration of intubation, expanding the sample size to 1889 patients. The results were similar for 30-day mortality and hospital LOS as regards effect size and statistical significance; however, the results were not statistically significant for 1-year mortality (*Table 2*). An additional sensitivity analysis excluded patients who died in the hospital- these results were consistent with the prior models and were statistically significant for hospital LOS, but not for mortality (*Table 2*).

As expected, the presence of nausea was not found to be associated with any impact on outcomes in the study cohort. As also would be expected, patients who had delirium had worse 30-day and 1-year mortality and longer hospital LOS.

## Discussion

We will first discuss the unexpected results that we observed in the data, and then go on to discuss the general issue of counterintuitive data. Our results indicating that increasing levels of patient-reported pain severity post-CABG surgery are associated with better clinical outcomes were not consistent with our initial hypothesis that better outcomes would correlate with better pain control as per the reported literature. In fact, prior studies have found increased levels of pain in the hospital to be associated with increased mortality. [11]

The difference in the study cohort between our study and others in the literature may explain the discordance in the findings. Our initial analysis was limited to "fast-tracked" patients after CABG: these are patients who did not have intra-operative complications and were extubated early in their ICU course. These patients made up 44% of the patients in the database and such patients are likely to be increasingly common as fast-track initiative are instituted. Studies that have reported worse clinical outcomes associated with post-operative pain did not select for a relatively healthy cohort of patients who had been screened and prepared for a successful CABG operation. Why would patients with higher levels of pain reported shortly after surgery have better outcomes? It is well documented that an increased inflammatory reaction is associated with increased pain. Pro-inflammatory cytokines such as IL-1β, IL-6, and TNF-α have been directly implicated in the physiology of pain.[12,13] These cytokines have also been found to be directly involved in the process of wound healing through the stimulation of various processes such as keratinocyte and fibroblast proliferation and synthesis and breakdown of extracellular matrix proteins.[14] We suspect that those patients who demonstrated better outcomes were mounting a more robust inflammatory response, leading to more pain, but importantly, also to an increased ability to heal.

Another possibility is that the higher pain levels detected represent a kind of proxy for a generally better state of health, including superior physiological function of the cardiovascular, respiratory, renal, and hepatic systems. In tandem, these systems act to metabolize and eliminate anesthetic and analgesic drugs so that the net pharmacokinetic result would likely be increased susceptibility to pain simply due to less of administered agents remaining at active sites. Furthermore, patients with better cardiovascular function would likely have better cerebral perfusion with improved central neurological function including pain perception, and thereby might have a pharmacodynamic reason for perceiving more pain, as well.  And of course, patients who are generally in better overall condition would, all else being roughly equal in terms of surgical issues, be expected to manifest better outcomes, as well. These thoughts are admittedly speculative and additional research is needed to explore these possibilities.

Clearly, it is critically important to point out that the goal of clinicians should not be in any way to maximize pain to optimize outcomes.  Conventional approaches that aim to control pain adequately should be employed.  Our observation is just that - an observation of an association and conjectures of possible linking mechanisms but is not intended in any way to drive pain management policy in the direction of tolerating undertreated pain.

We performed sensitivity analyses, one including all patients regardless of duration of mechanical ventilation post-operatively, and another excluding patients who died during hospitalization, and reached similar conclusions. When excluding in-hospital deaths, we discovered the 30-day mortality rate had a similar odds ratio but was no longer statistically significant. This is most likely due to the low mortality rate after hospital discharge following CABG, making it difficult to detect a statistically significant effect. This analysis would benefit from a repetition with additional data from a larger patient cohort.

Bias was limited as much as possible. We believe that researcher bias is a non-issue as these findings were not our expected results, but rather, the opposite. Sampling bias was also minimal. Our inclusion criteria were predefined prior to sampling the database. We also performed multiple sensitivity analyses to determine if those that were excluded would have had an effect on our results.  However, the study has several limitations inherent in any retrospective analysis of electronic health records. We recognize that correlation does not equal causation and further research is needed to determine the underlying physiologic mechanism for the results seen. Due to the self-reported nature of the pain scores, reporting bias is a concern. Some patients may have over-reported their pain while others under-reported their pain depending on their past experiences. We also recognize that the amount of pain medication administered is a confounder.

While we were unable to directly control for this due to lack of information regarding anesthetic and pain management in the database, we attempted to limit this potential confounder by excluding those with prolonged intubations who would inherently have received and required greater doses of sedatives and analgesics. Despite measures taken to guarantee internal validity, we anticipate appropriate skepticism with regard to generalizability of the findings beyond the highly selected group of patients who did well after CABG in the study. This, of course, is of genuine concern given the current state-of-affairs where clinicians are already inundated with conflicting studies of questionable quality. We therefore invite other investigators to replicate (and expand) our analysis in other databases.

As noted, our findings were contrary to clinical expectations and to most published works which associate increased pain with worse outcomes.[11,15-16] Encountering counterintuitive results is not unique to retrospective data analysis. Clinicians encounter unexpected, possibly aberrant values in a number of situations, especially in the evaluation of laboratory and device (e.g. monitor) data. When a possibly spurious lab result is obtained, the most frequent response is to repeat the test, as many of us were trained to do. Our responses to the repeated values would likely be of interest to behavioral economists who would spot confirmation bias when the second test came back with a more acceptable value and we then went about our business ignoring the unexpected value. But what if the second value comes back at the same or similar aberrant level? Do we repeat it again, or do we begin to believe that the value is 'real' and start to formulate a response to a potentially serious clinical problem? In this case, it is the *consistency* and *reproducibility* of the counterintuitive value that drives its possible validity, and the nature of this process is determined by the overall clinical danger (acuity, severity, treatability) of that value being truly real. The consistency we found in the pain score values also drove us to consider the possibility that the values were 'real' even though they were counterintuitive in terms of our expectations. Since we were analyzing data retrospectively, we were not in a position to attempt to reproduce the values, but this is a possible target for future research.

Another factor that will drive the approach to counterintuitive values is whether they are *possible*. Impossible values would include items such as a potassium of 64.5 or a CVP of 700- these are simply impossible values. But a potassium of 7.3 or a CVP of 28 are both possible (and very worrisome) values. The pain values we obtained in the better outcome were higher than expected, but not so high that we would expect these patients to be screaming in agony.  In other words, while unexpected and perhaps unusual, they were not impossible in the context in which we observed them.

One question that would arise with a potassium of 7.3 would be that of continuity- did the value occur suddenly in the context of normal values, or did it occur in the context of gradually or even consistently abnormal values?  In the context of persistently very abnormal values, e.g. untreated uremia, it would be the normal value that might appear counterintuitive.  So that while most counterintuitive values will tend to be out of the 'normal range', they will not necessarily be so.  In the context of increasing values, it might simply be the first one that was not only out of the range of normal values, but had crossed the line into what might be considered the range of critical values,

The question arises whether counterintuitive results are actually false, and therefore misleading, results.  Table Three shows a categorization of the types of errors that could result in faulty data.  Human error has always been a possibility in data entry but may even be exacerbated by problems at the human- computer interface. It is unlikely that the entries we observed were due to a fundamental misunderstanding of the pain scale as they were entered by multiple users over a prolonged period of time. Our particular example was not one of laboratory or device error, but these are likely the most common sources of what appear to be faulty data for the reasons listed in the table. Once the human (or interface engine) has entered the data in the software, a number of problems can then intervene including software, hardware, systems, and data analytic errors. We are not able to attribute the counterintuitive data we observed to any of these factors, however.

There are multiple steps that should be taken following the detection of counterintuitive results in retrospective data analysis. While able three displays the issues to be addressed,  Table four displays a suggested organization of the characterization of reliable vs faulty data that should be kept in mind when confronted with counterintuitive data. The first step is to retrace the process and workflow involved in data entry so far as this is possible. The data we studied was obtained at the institution of several of the authors where nurses are trained to assess pain on a standard scale from 0 to 10. There are several potential faults to this method. The nursing staff could neglect to regularly assess pain or they may neglect to enter the information into the medical record from which the database is derived. While this may alter a few data points, it is unlikely to systematically affect all the data unless there was an obvious glaring institutional issue affecting every nurse and every data entry.

After determining that the data source is sound and valid, additional statistical tests can be run on the patient cohort. Tests, such as falsification hypothesis testing as demonstrated in the case, can add validity to the results if it can be shown that the cohort follows other generally known principles. In the case of falsification hypothesis testing, one shows that factors with no known effect on the outcomes of interest do not have an association in the analytic cohort. In our case we used nausea, which is a common side effect experienced by many in hospital patients that does not have any known effect on hospital LOS or mortality. In some cases, one could also employ concurrent contextual data to help confirm the veracity of data- for example, one could examine ECG tracings if hyperkalemia was the counterintuitive result being analyzed. In order to do this, we examined concomitant vital signs during the time of pain measurements. We would have liked to have observed significant increases in heart rate, respiratory rate, and blood pressure associated with higher pain levels, but in fact, we did not:With the combination of the use of analgesics, residual anesthetic agents, and the concurrent use of drugs that directly affect vital signs such as beta-blockers, the lack of correlation is probably not surprising. In this setting, it appears to be inadvisable to use vital sign changes as a proxy for the presence of unvoiced pain. Finally, one can attempt to physiologically explain the disparity between the observed and expected results as we did above for the case of post- CABG pain being greater in more vigorous patients with more robust metabolic capabilities.

In addition to the issues mentioned in the introduction, the use of lower thresholds for blood transfusions in the ICU is an example of a counterintuitive finding. Prior to this work, ICU target hemoglobin and hematocrit levels were set at greater than 10 g/dL and 30% respectively, theoretically to ensure adequate oxygen delivery.[17] This led to increased transmission of blood borne diseases, unnecessary healthcare expenditures, and actually worse outcomes.[18] Data later showed that this 10/30 rule was not necessary for most patients, but only for selected patients such as those with acute coronary syndrome whose hemoglobin is less than 10g/dL and actively experiencing chest pain. The initially counterintuitive findings that lower hemoglobin levels were not only acceptable but preferable in most cases, served as research triggers to more fully elucidate what was truly occurring in this context. Our case may serve as an analogous research trigger in terms of optimally managing postoperative pain. Outcomes such as mortality and LOS are complex phenomena driven by many factors- to observe a clear and robust statistical effect such as we observed is strongly suggestive that something 'real' is occurring even if the data are counterintuitive and we don't understand all the myriad factors that may have contributed to the outcomes observed.

The final step when dealing with counterintuitive data is to look for additional evidence that confirms the reliability of the results (perhaps this could be termed 'confirmatory metadata'). With respect to our CABG case, the analysis should be rerun on additional databases and in different settings to confirm or refute our initial results. Just as clinicians continued to manage intensive care unit anemia as they always had until more definitive results were reported, our results should not impact the analgesic care of patients at this point. However, we hope that we have raised the issue in the appropriate minds that outcomes may benefit from approaches slightly different from usual. After all, one can easily eliminate all pain from postoperative patients but they would have to remain sedated and ventilated for an indefinite period of time to do so. And after they are extubated, pain management should not be so aggressive that it leads to apnea and respiratory arrest. In other words, there may be a mild level of tolerable pain that leads patients to their best outcomes, and no honest clinician will guarantee a patient that they will have no pain at all after a procedure like a sternal-disrupting CABG.

## Conclusion

Contrary to our expectations, we observed, in a retrospective analysis of electronic health records, that post-CABG fast-track patients with higher pain scores had better outcomes. The increasing use of EHRs for secondary analysis will likely lead to an increasing incidence of such apparently counterintuitive results. While the first step in this situation is to attempt to confirm the reliability of both the analytic process and the data itself, such findings that prove to be robust may lead to further ideas and subsequent research that drive future clinical care. On the other hand, clinicians must be careful in terms of modifying their practices until the implications of such counterintuitive (or any) data have been thoroughly vetted and confirmed in diverse database contexts and via the peer review process.

Effect of postoperative epidural analgesia on morbidity and mortality following surgery in medicare patients. Regional Anesthesia and Pain Medicine, 29(6), 525-533. doi:10.1016/j.rapm.2004.07.002

[17] Wang J, Klein H. Red blood cell transfusion in the treatment and management of anaemia: the search for the elusive transfusion trigger. Vox Sanguinis [serial online]. January 2010;98(1):2-11. Available from: Academic Search Complete, Ipswich, MA. Accessed June 8, 2018.

[18] Perioperative Red Blood Cell Transfusion. *JAMA*. 1988;260(18):2700–2703. doi:10.1001/jama.1988.03410180108040

# Figures

**Figure 1:** Patient Cohort selection

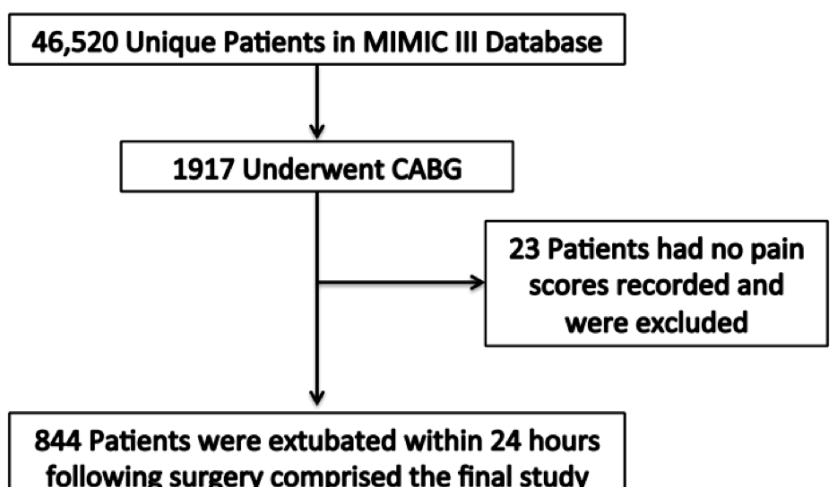

FIGURE 1: SHOWS SELECTION OF PATIENT COHORT FROM MIMIC DATABASE. AFTER SELECTING THOSE WHO UNDERWENT CABG PROCEDURE AND EXCLUDING THOSE WITH NO PAIN MEASUREMENTS; 844 PATIENTS WERE EXTUBATED WITHIN 24 HOURS FOLLOWING SURGERY AND INCLUDED IN THE COHORT.

**Figure 2:** Bivariate Analysis

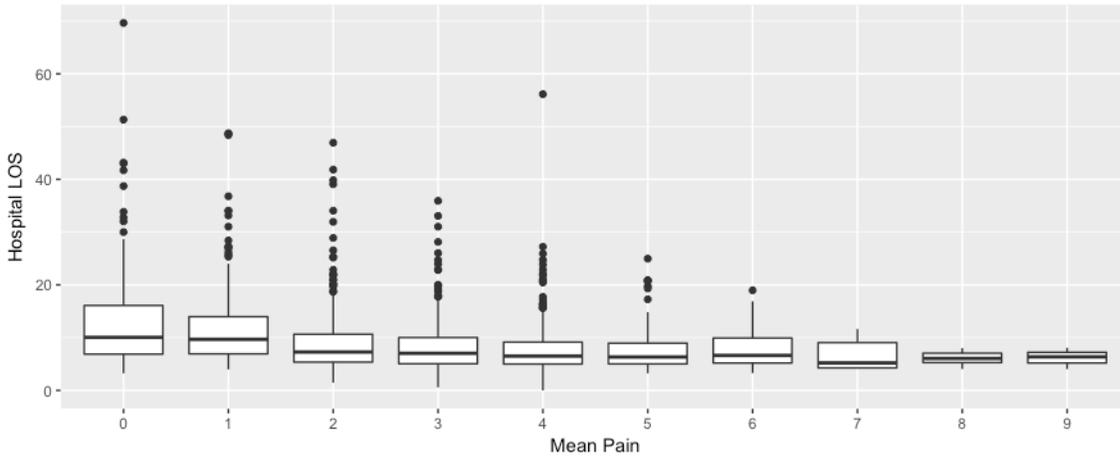
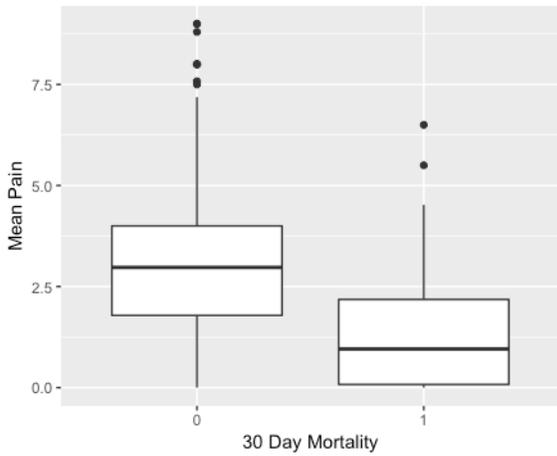
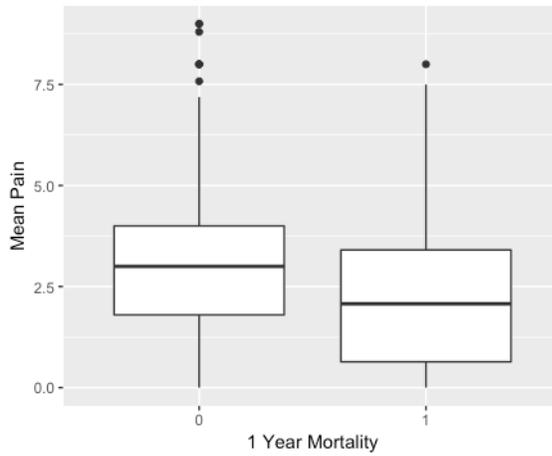

**FIGURE 2: THREE PLOTS DEMONSTRATING THE BIVARIATE RELATIONSHIP BETWEEN THE OUTCOMES OF INTEREST AND MEAN PAIN. PLOT A SHOWS DECREASED LENGTH OF STAYS WITH INCREASED MEAN PAIN LEVELS. PLOT B AND PLOT C SHOW THAT, ON AVERAGE, THOSE WHO EXPIRED AT 30 DAYS AND 1 YEAR MARKS EXPERIENCED LOWER IN HOSPITAL PAIN LEVELS THAN THOSE WHO DID NOT EXPIRE.**

# Tables
**Table 1:** Cohort Characteristics

TABLE 1: SHOWS THE DISTRIBUTION OF THE OUTCOMES AND COVARIATES IN THE PATIENT COHORT.

ABBREVIATIONS: OASIS = OXFORD ACUTE SEVERITY OF ILLNESS SCORE; E_SCORE = ELIXHAUSER SCORE.

OASIS SCORES RANGE FROM 0 TO 75, WITH HIGHER SCORES INDICATING MORE SEVERE DISEASE;

ELIXHAUSER SCORE WAS BASED ON AGGREGATE PRESENCE OR ABSENCE OF CONDITIONS IN ELIXHAUSER COMORBIDITY INDEX. COUNTS RANGED FROM 0 TO 9.

|  |  | No Pain | Mild | Moderate | Severe | p |
|---|---|---|---|---|---|---|
| n |  | 68 | 419 | 336 | 21 |  |
| Age (mean (sd)) |  | 71.50 (10.61) | 67.75 (10.54) | 64.98 (9.73) | 65.13 (12.85) | <0.001 |
| Gender = male |  | 45 (66.2) | 333 (79.5) | 282 (83.9) | 14 (66.7) | 0.003 |
| oasis (mean (sd)) |  | 31.96 (7.25) | 30.32 (6.47) | 31.44 (6.35) | 30.57 (6.20) | 0.056 |
| e_score (%) |  |  |  |  |  | <0.001 |
|  | 0 | 4 (5.9) | 96 (22.9) | 87 (25.9) | 7 (33.3) |  |
|  | 1 | 12 (17.6) | 116 (27.7) | 97 (28.9) | 4 (19.0) |  |
|  | 2 | 12 (17.6) | 81 (19.3) | 79 (23.5) | 4 (19.0) |  |
|  | 3 | 10 (14.7) | 61 (14.6) | 46 (13.7) | 3 (14.3) |  |
|  | 4 | 12 (17.6) | 29 (6.9) | 16 (4.8) | 1 (4.8) |  |
|  | 5 | 6 (8.8) | 19 (4.5) | 8 (2.4) | 2 (9.5) |  |
|  | 6 | 7 (10.3) | 8 (1.9) | 2 (0.6) | 0 (0.0) |  |
|  | 7 | 2 (2.9) | 4 (1.0) | 1 (0.3) | 0 (0.0) |  |
|  | 8 | 0 (0.0) | 4 (1.0) | 0 (0.0) | 0 (0.0) |  |
|  | 9 | 3 (4.4) | 1 (0.2) | 0 (0.0) | 0 (0.0) |  |
| Mortality |  |  |  |  |  |  |
| In Hospital |  | 9 (13.2) | 5 (1.2) | 1 (0.3) | 0 (0.0) | <0.001 |
| 30 Day |  | 10 (14.7) | 10 (2.4) | 1 (0.3) | 0 (0.0) | <0.001 |
| 1 Year |  | 16 (23.5) | 22 (5.3) | 7 (2.1) | 1 (4.8) | <0.001 |

**Table 2:** Primary Outcome Results

TABLE 2: SHOWS RESULTS FROM MAIN ANALYSIS AND THE TWO SENSITIVITY ANALYSES.
*, **, *** DENOTES SIGNIFICANCE AT THE 90%, 95%, AND 99% LEVEL, RESPECTIVELY.

| Model | 30 Day Mortality Odds (95% Confidence Interval) | 1 Year Mortality Odds (95% Confidence Interval) | Length of Stay Estimate |
|---|---|---|---|
| Primary Analysis: | | | |
| Mean Pain | 0.457*** (0.304 – 0.687) | 0.710*** (0.571 - 0.881) | -0.916*** |
| Median Pain | 0.639*** (0.466 - 0.877) | 0.856* (0.727 - 1.008) | -0.696*** |
| Max Pain | 0.812*** (0.693 - 0.951) | 0.887** (0.790 - 0.995) | 0.148* |
| Categorical Pain | 0.214*** (0.091 - 0.502) | 0.450*** (0.266 - 0.760) | -2.270*** |
| Sensitivity Analysis 1: Including all patients regardless of intubation length | | | |
| Mean Pain | 0.592*** (0.456 - 0.768) | 0.898 (0.785 - 1.027) | -0.709*** |
| Categorical Pain | 0.328*** (0.184 - 0.586) | 0.740* (0.527 - 1.037) | -1.706*** |
| Sensitivity Analysis 2: Excluding hospital mortality patients | | | |
| Mean Pain | 0.803 (0.567 - 1.137) | 1.027 (0.889 - 1.187) | -0.701*** |
| Categorical Pain | 0.709 (0.309 - 1.625) | 1.038 (0.714 - 1.509) | -1.680*** |

**Table 3:** Putative Reasons for Truly Faulty Data

| | |
|---|---|
| Human error | Mis-entry; misunderstanding of scale values; faulty understanding of use of data entry software; faulty interpretation of device values |
| Lab error | Sampling error (eg hemolysis); measurement error |
| Device error | Disconnect, interference, faulty calibration, software error; unexplained, transient aberrant values that resolve and do not recur |
| Systems error | Interface error, application interoperability error |
| Software error | Bug in software relating to data value entry; data wrongly captured, stored, and/or retrieved due to software design faults or bugs |
| Hardware error | Hardware issues that impact software and systems |
| Data analytic error | Error in analytic algorithm or process |

**Table 4:** Criteria to establish possible validity of counterintuitive data

| | |
|---|---|
| Viability | Is the value consistent with clinical reality? Are the values even possible ones? |
| Consistency | If applicable (not always the case in retrospective analysis), is the value observed consistently, such as in our pain score observations? |
| Continuity | What is the context of the value- does it occur as a sudden aberrant value, or as one of increasingly aberrant values? |
| Identity | Are the circumstances that produced the data truly identical so far as identifiable? Ie. Would the same circumstances produce the same data results in a different database, institutional, or cultural context? |

| | | |
|---|---|---|
| | Reproducibility | Is the value reproducible on repetition? while reproduction can not be performed in retrospective data, can it be reproduced upon observation across different clinical databases |
| | Sensibility | Even if it does not meet current clinical expectations and is possible, does it make potential sense in associated clinical context? |
| | Curiosity | Does it drive the observer to seek alternative better solutions and pose questions for further research? |